\documentclass[twocolumn,showpacs,preprintnumbers,amsmath,amssymb]{revtex4}
\usepackage{tabularx,graphicx}\begin{document}
%\documentstyle[aps]{revtex}
%\documentstyle[preprint,aps]{revtex}
%\begin{document}
\newcommand{\beq}{\begin{equation}}
\newcommand{\eeq}{\end{equation}}
\newcommand{\beqn}{\begin{eqnarray}}
\newcommand{\eeqn}{\end{eqnarray}}
\newcommand{\bmath}{\begin{subequations}}
\newcommand{\emath}{\end{subequations}}
%\draft
\title{Comment on Theory of spin Hall effect}
\author{J. E. Hirsch}
\address{Department of Physics, University of California, San Diego\\
La Jolla, CA 92093-0319}

\begin{abstract} 

Chudnovsky\cite{chud} recently proposed (arXiv: 0709.0725) a beautifully simple mechanism for the intrinsic spin Hall effect involving an extension 
of the Drude model. 
We point out that an equivalent proposal was made by us in an alternative treatment of the problem  
in Phys.Rev. {\bf B60}, 14787 (1999)\cite{jh}, reaching identical conclusions.
The coincidence of both approaches supports their physical validity.
The mechanism applies to the spin Hall effect in all metals and semiconductors and to the anomalous Hall effect in all ferromagnetic
metals and semiconductors.

\end{abstract}
\maketitle

The theoretical study of spin transport in solids in connection with the spin Hall effect\cite{sh1,sh2} and the anomalous Hall effect of
ferromagnetic metals\cite{ah} is mired in controversy. Since the original work of Karplus and Luttinger\cite{karp} and its
refutation by Smit\cite{smit}, no consensus has  been reached on the fundamental question whether or not transverse spin transport
can be induced by a perfectly periodic lattice potential  in the absence of impurity scattering\cite{sinova}.

Chudnovsky has recently shown\cite{chud} that incorporating the spin-orbit interaction in the single particle Hamiltonian and considering the
equations of motion within a simple Drude treatment results in a {\it  spin-dependent transverse velocity}
\beq
<\dot{\vec{r_1}}>=\frac{\hbar e^2 \tau^2 A}{4m_e^3c^2}\vec{\sigma} \times \vec{E}
\eeq
where $\vec{E}$ is the applied longitudinal electric field, $\tau$ is the relaxation time, $m_e$ the electron mass,
and $A$ an average of the second derivatives of the crystal potential over the unit cell volume, given in particular for a 
cubic crystal by

\beq
A=<\frac{\partial^2\phi}{\partial x^2}>=\frac{4\pi}{3} \frac{Q}{a^3}
\eeq
where $\phi(x,y,z)$ is the periodic lattice potential resulting from ionic charges $Q$ on a simple cubic lattice of
side length $a$. 

In Phys.Rev.{\bf B60}, 14787 (1999)\cite{jh} we examined the force exerted on a moving spin by the crystal potential by consideration of the 
equivalent electric dipole\cite{boyer}
\beq
\vec{p}=\gamma \frac{\vec{v}}{c} \times\vec{m}
\eeq
where $\vec{m}$ is the magnetic moment 
($\vec{m}=\mu_B \vec{\sigma}$)  propagating with drift velocity $\vec{v}$, and $\gamma = ( 1 - v^2/c^2 ) ^{-1/2} \sim 1$. 
This electric dipole interacts electrostatically with the local electric field $\vec{E_c}$ originating in the crystal potential and
experiences a force proportional to the electric field gradient
\beq
\vec{F}_t=\vec{\nabla}({\vec{p}\cdot\vec{E_c}}).
\eeq
The magnitude and direction of this force
varies with the position in the unit cell (Fig. 2 of ref.\cite{jh}).
Upon averaging over particle trajectories\cite{ave}
we found a net force antiparallel to the electric dipole (i.e. perpendicular to the spin and propagation directions)
given by
(Eq. 16) of ref.\cite{jh})
\beq
<\vec{F}_t>=-\frac{2\pi}{3} \frac{Q}{a^3} \vec{p}.
\eeq
It is easily seen that Chudnovsky's result Eq. (1) is equivalent to Eq. (5), i.e. 
\beq
\frac{m_e<\dot{\vec{r_1}}>}{\tau}=<\vec{F}_t>
\eeq
with $A$ given by Eq. (2) and the drift velocity $\vec{v}=(e\tau/m_e)\vec{E}$.

Eq. (5) can be understood as an 'effective' magnetic field $B_{eff}$   deflecting the propagating electron just as an applied
magnetic field does in the ordinary Hall effect 
($\vec{B}_{eff}=(2\pi/3)(Q/(a^3|e|))\vec{m}$).  The effect is strong: for example, for a monovalent cubic crystal with
 $a=1\AA$, Eq. (5) corresponds to an effective magnetic field $B_{eff}=1.94 T$.

In our original proposal of the spin Hall effect\cite{sh2}, we pointed out that it would result from the same mechanism(s) giving rise
to the anomalous Hall effect of ferromagnetic metals, and that in particular
it would arise from the intrinsic effect discussed here (Ref.3 of ref.\cite{sh2}).
 Posteriorly other different mechanisms for intrinsic spin Hall effect 
were proposed by Murakami et al\cite{zhang} and by Sinova et al\cite{sinova2}
 and attracted a lot of attention, applicable to particular lattice structures (e.g. lacking inversion symmetry), band structures and
Hamiltonians\cite{sinova}.
In contrast, the mechanism discussed here applies to $all$ metals and semiconductors.

\acknowledgements
The author is grateful to E.M. Chudnovsky for suggesting that I write this comment.


\begin{references}
\bibitem{ave} Note that in contrast to Chudnovsky we average over particle trajectories rather than over the
unit cell volume. Trajectories do not go through lattice points.
\end{references}
\end{document}